\documentclass[twocolumn,superscriptaddress, preprintnumbers,prb]{revtex4-1}
\usepackage{graphicx}
\usepackage{dcolumn}
\usepackage{amssymb}
\usepackage{amsmath}
\usepackage{float}
\usepackage{setspace}
\usepackage{hyperref}
\usepackage{url}
\usepackage{soul}
\usepackage{lipsum}

\hypersetup
{
colorlinks=true, 
citecolor=blue
}

\begin{document}

\title{Nonlinear topological laser based on multipole insulators}

\author{Zi-Yuan Li}
\affiliation{School of Electronic Engineering, North China University of Water Resources and Electric Power, Zhengzhou 450046, People's Republic of China}
\affiliation{Lanzhou Center for Theoretical Physics, Key Laboratory of Theoretical Physics of Gansu Province, Lanzhou University, Lanzhou 730000, China}

\author{Zi-Xiang Hu}
\affiliation{Department of Physics and Chongqing Key Laboratory for Strongly Coupled Physics, Chongqing University, Chongqing 401331, People's Republic of China}

\author{Qi Li}
\email{liqicqu@126.com}
\affiliation{GBA Branch of Aerospace Information Research Institute, Chinese Academy of Sciences, Guangzhou 510700, China}
\affiliation{Thrust of Advanced Materials, The Hong Kong University of Science and Technology (Guangzhou), Nansha, Guangzhou 511400, China}

\date{today}
\begin{abstract}
Two-dimensional higher-order topological insulators (HOTIs), characterized by distinctive one-dimensional edge states and zero-dimensional corner states, provide an ideal platform for developing higher-order topological lasers. In this work, we systematically investigate the two-dimensional Benalcazar-Bernevig-Hughes (BBH) model, which hosts quantized quadrupole moments and topologically protected corner and edge states. By confining the lasing mode to selected topological corner or edge states under controlled gain, we demonstrate that the stable light excitation achieved after long-time evolution is predominantly determined by the topological properties of the model Hamiltonian.
To characterize the system's topological features, we introduce several diagnostic ratios: the corner decay ratio $\tau_{1}$ and edge-to-corner ratio $\tau_{2}$ quantify the localization degree and spatial extent of corner states, respectively, while the inter-corner transfer ratio $\chi$ measures the intensity transfer efficiency mediated by coherent edge-state dynamics. The abrupt changes in $\tau_{1}$ and 
$\tau_{2}$ as functions of the hopping parameter $\gamma/\lambda$ directly reveal topological phase transitions, providing a comprehensive toolkit for extracting topological signatures from the system's dynamical evolution. Additionally, modulating the lattice site parity enables flexible tuning of corner state localization positions, offering insights for device engineering. Our calculations reveal that achieving bistability between corner states and edge states is relatively challenging.
\end{abstract}

\maketitle
\section{introduction}
Topological physics initially studies how material geometry governs electronic properties, leading to robust edge state‌s and corner states ~\cite{Hasan,QiXL,Das,ShenSQ}.‌ The same mathematical framework that describes topologically protected electronic states can be adapted to other wave phenomena, including electromagnetic waves in optical systems. By engineering artificial lattices and photonic crystals with specific geometric configurations, researchers can create optical analogues of topological insulators and other exotic electronic phases~\cite{lu2014topological}.

Nanoscale photonic chips are essential for modern optics, but fabrication defects and impurities often degrade their performance by causing optical losses. Topological photonics offers a solution by creating edge and corner states that are inherently immune to defects and perturbations~\cite{WangNature,shalaev2019robust}. 
The integration of this approach with semiconductor lasers allows topological lasers to achieve robust, single-mode, and low-threshold operation with efficient light emission \cite{ZhaoNc,PartoPRL,XuXiulaiLight,kim2020multipolar}. The fundamental principle of topological lasers relies on topologically protected optical modes, which are determined by the topological properties of the dielectric structure and achieve lasing through optical pumping \cite{ozawa2019topological,SegevScienceTheo,SegevScienceExpe}. These modes fall into two main categories. The first is topological edge states—one-dimensional modes at interfaces between distinct topological phases—that enable unidirectional
light transport due to their robustness against defects and perturbations~\cite{PartoPRL, JeanNP}. 
The second type consists of topological corner states, which are zero-dimensional modes localized at the corners of two or three-dimensional topological insulators. Compared to edge states, corner states offer stronger spatial confinement, smaller mode volume, and significantly higher quality factors (Q-factors)~\cite{bahari2017nonreciprocal}. As a result, lasers based on topological corner states exhibit higher efficiency and lower lasing thresholds~\cite{XuXiulaiLight}.

Building on these principles, experimental realizations of topological lasers have been demonstrated through distinct models with unique structural and performance features. The microscale topological lasers, based on the  Su-Schrieffer-Heeger (SSH) model, use micropillar or microring arrays with InGaAsP/InP quantum wells as gain medium~\cite{PartoPRL,ZhaoNc}. By modulating coupling strengths in dimerized chains, they create topological band gaps and edge states, enabling single-mode lasing robust against perturbations and validating chiral symmetry protection. Using photonic crystal nanocavities, people achieve strong light-matter interaction for the laser with a mode volume of $0.23(\lambda/n)^3$, quality factor of 59,700, and low threshold of 46 $\mu W$~\cite{Ota}. Key parameters are systematically tunable while preserving single-mode operation.

The current frontier in topological laser research is centered on higher-order topological insulators (HTOIs), which extend the bulk-boundary correspondence by allowing topological invariants (multipole moments) to be defined on the boundaries themselves~\cite{schindler2018higher, khalaf2018higher,BernevigScience, BernevigPRB}. This enables topological robustness to lower-dimensional states, such as corner or hinge modes. Deep reinforcement learning is used to efficiently perform inverse design of high-Q nanocavities supporting topological corner states ~\cite{yu2025inverse,sun2026ai}. Engineered artificial structures provide highly tunable platforms for realizing HOTIs, particularly in photonic systems where coupled resonators and photonic crystals enable precise control over light~\cite{PengYang, FangChen,zhang2025topological,wu2026reconfigurable}. These implementations are guided by theoretical frameworks such as quantized multipole insulators, generalized SSH models, and breathing kagome lattices, demonstrating that higher-order topology is accessible through electromagnetic modalities. In surface-wave photonic crystals, HOTIs support robust cavity modes suitable for compact photonic devices ~\cite{ZhangAdvancedScience}. Experimental realizations in two-dimensional dielectric photonic crystals based on the SSH model have visualized topological edge and corner states using near-field scanning techniques ~\cite{ChenYanfengPRL,ChenYanfengPRB, MarcoNaturePhysics}. Higher-order topological insulators such as breathing kagome and pyrochlore lattices exhibit fractional corner charges characterized by quantized polarizations, and their quench dynamics enable real-time observation of topological state evolution ~\cite{EzawaPRR,EzawaPRL,EzawaPRB}.

Here we focus on the 2D BBH model, which hosts a quantized quadrupole moment $q_{xy}=\pm{e/2}$ and exhibits topologically non-trivial corner states with fractional charge ~\cite{BernevigScience,BernevigPRB}. Unlike the SSH chain and breathing kagome lattice, the BBH model achieves topological protection through quantized multipole moments protected by Berry-phase topology and non-commuting reflection symmetries $\{M_{x},M_{y}\}=0$, resulting in four corner states with fractional charge $\pm{e/2}$. In contrast, corner/edge state protection in the SSH chain and breathing kagome lattice is rooted in different mechanisms. The SSH chain's end states are protected by the Zak phase under chiral symmetry, while breathing kagome corner states originate from SSH-type topological conditions along the two directions meeting at a corner. Through destructive interference and discrete symmetry constraints, these systems produce exactly solvable zero-energy corner modes—two for the 1D SSH and three for the triangular kagome—within their respective nontrivial topological phases \cite{BergholtzPRB18}. Both SSH and breathing kagome lattices are symmetry-protected systems where corner/edge states persist due to the interplay between intrinsic bulk topology and lattice symmetries.

In this paper, we systematically compare the corner intensity distributions before and after dynamical revolutions in the BBH systems, revealing that the spatial confinement pattern remains qualitatively invariant throughout the lasing dynamics—a striking consistency indicating that corner state topology dictates the laser emission profile. Numerical simulations further demonstrate that the characteristics of spatially-localized intensity distribution serve as a direct diagnostic for identifying the parameter regimes hosting non-trivial topological corner states. To extract the characteristic topological signatures from the real-time lasing dynamics, we define two quantifiable observables $\tau$ which tracks the corner intensity buildup and effectively maps out the boundary of the non-trivial topological phase, and $\chi$ which characterizes the spatial spreading of laser intensity along the edge. Together, these metrics provide a comprehensive description of how topological protection manifests dynamically during the lasing process. 

\section{model and method}
 Topological lasers can be realized through arrays of coupled, active resonators, which serve as the fundamental building blocks providing the necessary optical gain. These systems are inherently non-Hermitian, incorporating both gain and loss. To specifically promote the lasing of topological edge modes, gain is typically provided to  specific regions (edge or corner). The fundamental gain mechanism in topological insulator lasers relies on the interplay between saturable gain and topological protection. This character is crucial for ensuring stable laser operation.
We applied such a model to describe the lasing dynamics
\begin{equation}
    i\frac{\partial \Psi_{n}}{\partial t} = H \Psi_{n} -i \alpha (1 - \frac{\xi P_n}{1+\vert \Psi_{n} \vert^2/\eta})\Psi_{n}
    \label{Eq1}
\end{equation}
where $H$ represents target Hamiltonian which hosts non-trivial topological order, $\Psi_{n}$ is the wavefunction of $n$th site in lattice, $\alpha$ denotes the background loss, $\alpha\xi$ introduces the nonlinear saturable gain in the stimulated sites, $\eta$ accounts for the self-saturation nonlinearity.  
The value of $P_{n}$ is conditionally assigned: it takes the value of 1 for corner or edge positions, and 0 for all other sites.
\begin{figure}[htbp]
\centering
\includegraphics[width=9cm]{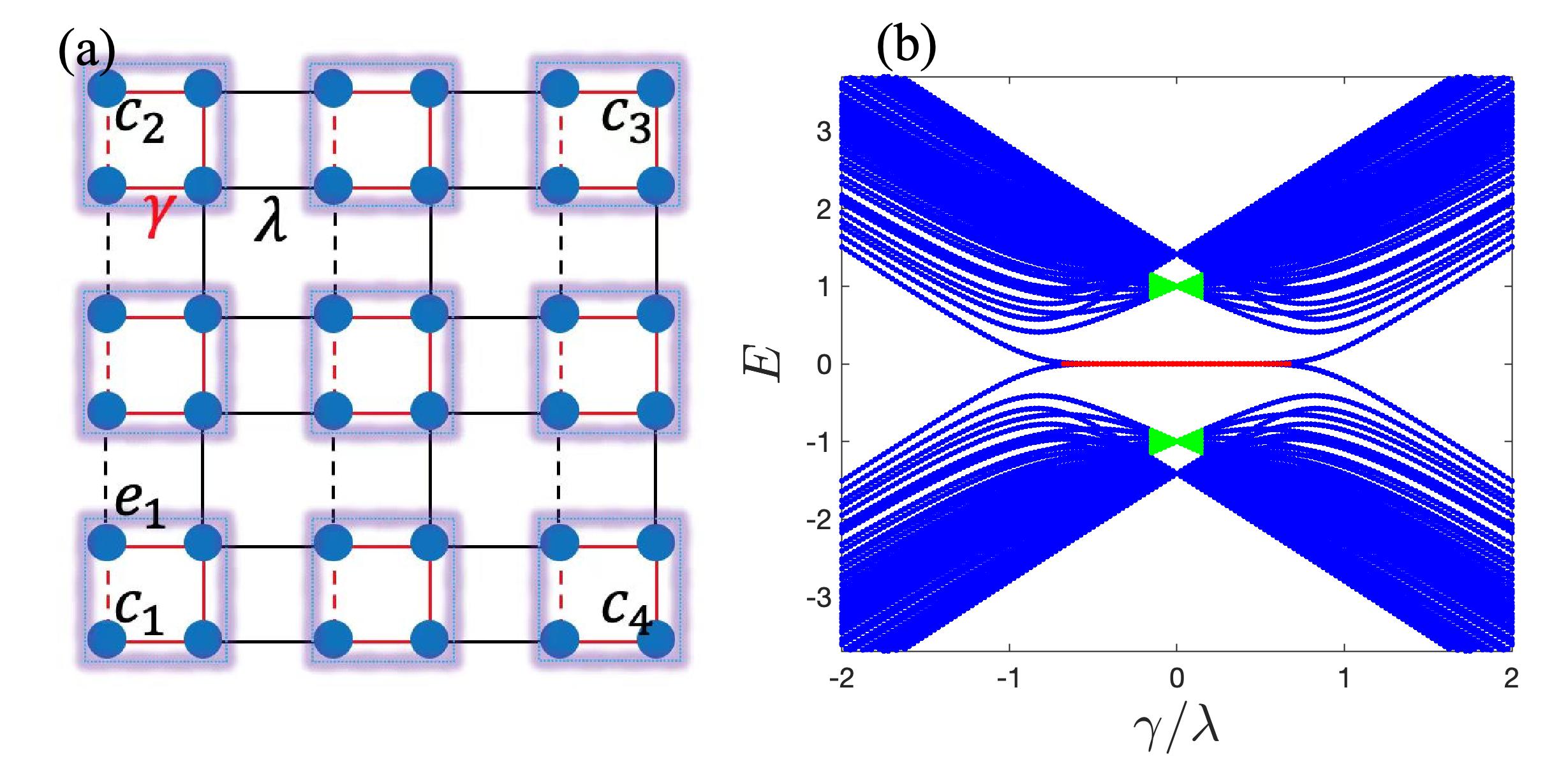}
\caption{(a) Sketch of the BBH model. $\gamma$, $\lambda$ represent intra- and inter- hopping strengths, and dashed lines represent hopping terms with negative signs. $c_i$ ($i=1-4$) represent four corner sites and $e_1$ stands for one edge site.  (b) Energy spectrum as a function of $\gamma/\lambda$ for BBH lattice with $20 \times 20$ sites, where the in-gap edge and corner states are marked in green and red, respectively.}
\label{BBH-sketch}
\end{figure}

The target Hamiltonian $H$ represents paradigmatic models exhibiting topological features that originate from alternating couplings, allowing for their flexible application across diverse physical contexts. To realize HOTIs with robust, disorder-immune operation, the choice of target Hamiltonian is governed by the topological band structure quantization, physical implementation feasibility, and lasing performance optimization. For instance, the Haldane model emerges as the primary architecture because it breaks time-reversal symmetry through staggered magnetic flux ($\phi=\pi$ per plaquette), generating quantized Chern numbers ($ \mathcal{C}=\pm 1$) that guarantee unidirectional chiral edge states immune to backscattering, that essential for single-mode lasing without magnetic field requirements \cite{SegevScienceTheo}. Experimentally it employs aperiodic coupler arrays creating artificial gauge fields through spatially-shifted intermediary links with accumulated hopping phases ( $\pm2\pi\alpha $ per unit cell), achieving analogous topological non-triviality without exotic materials \cite{SegevScienceExpe}.

Our study focuses on the lasing dynamics driven by these non-trivial topological properties in BBH model. Its 4-band Bloch Hamiltonian in momentum space can be written as
\begin{equation}
	\begin{split}
		H_{BBH}& = [\gamma + \lambda \cos(k_{x}) ]\Gamma_4 + \lambda \sin(k_{x}) \Gamma_3 \\ &+ [ \gamma + \lambda \cos(k_y) ] \Gamma_2 + \lambda \sin(k_y) \Gamma_1 ,
	\end{split}
\end{equation}
that $\gamma$, $\lambda$ represent intra- and inter-hopping strengths, $\Gamma_{0}=\tau_{3}\sigma_{0}$, $\Gamma_{k}=-\tau_{2}\sigma_{k}$ and $\Gamma_{4}=\tau_{1}\sigma_{0}$, for $k=1,2,3$ and $\tau$, $\sigma$ are Pauli matrices.
The parallel tight-binding model, illustrated in Fig.~\ref{BBH-sketch}(a), describes spinless electrons on a square lattice and bond dimerization along the x and y directions. Each unit cell contains four sites, indicated in purple box and four corner sites are labeled as $c_i$. 

To realize the topological laser based on the BBH model, the laser-written photonic crystals in fused silica is proposed \cite{BernevigScience}. This approach overcomes the prior challenge of requiring hoppings with opposite signs by introducing a negative coupler construction that inserting an auxiliary waveguide to produce effective negative hopping\cite{keil2016prl}.

\section{numerical results}
\subsection{ Corner state characteristics and lasing dynamics in even-sized 2D BBH lattice}
We start to investigate the lasing dynamics and topological properties of the two-dimensional BBH model, characterized by the presence of quantized corner states and in-gap edge states. Through the examination of the energy spectrum and intensity distribution under varying hopping parameters, we unveil the emergence of localized states and their response to external stimulation across different topological phases. 
Fig.~\ref{BBH-sketch}(b) depicts the energy spectrum for the case of a finite $ N \times N$  BBH lattice, with $N$  being even. In-gap edge states emerge within the parameter range $|\gamma/\lambda|<0.18$ for $N=20$, exhibiting $(N-2)\times2$ degeneracy in both positive and negative energy bands. Meanwhile, corner-localized states demonstrate four-fold degeneracy at zero energy in parameter space $|\gamma/\lambda| \le 1$.

\begin{figure}[htbp]
\centering
\includegraphics[width=9cm]{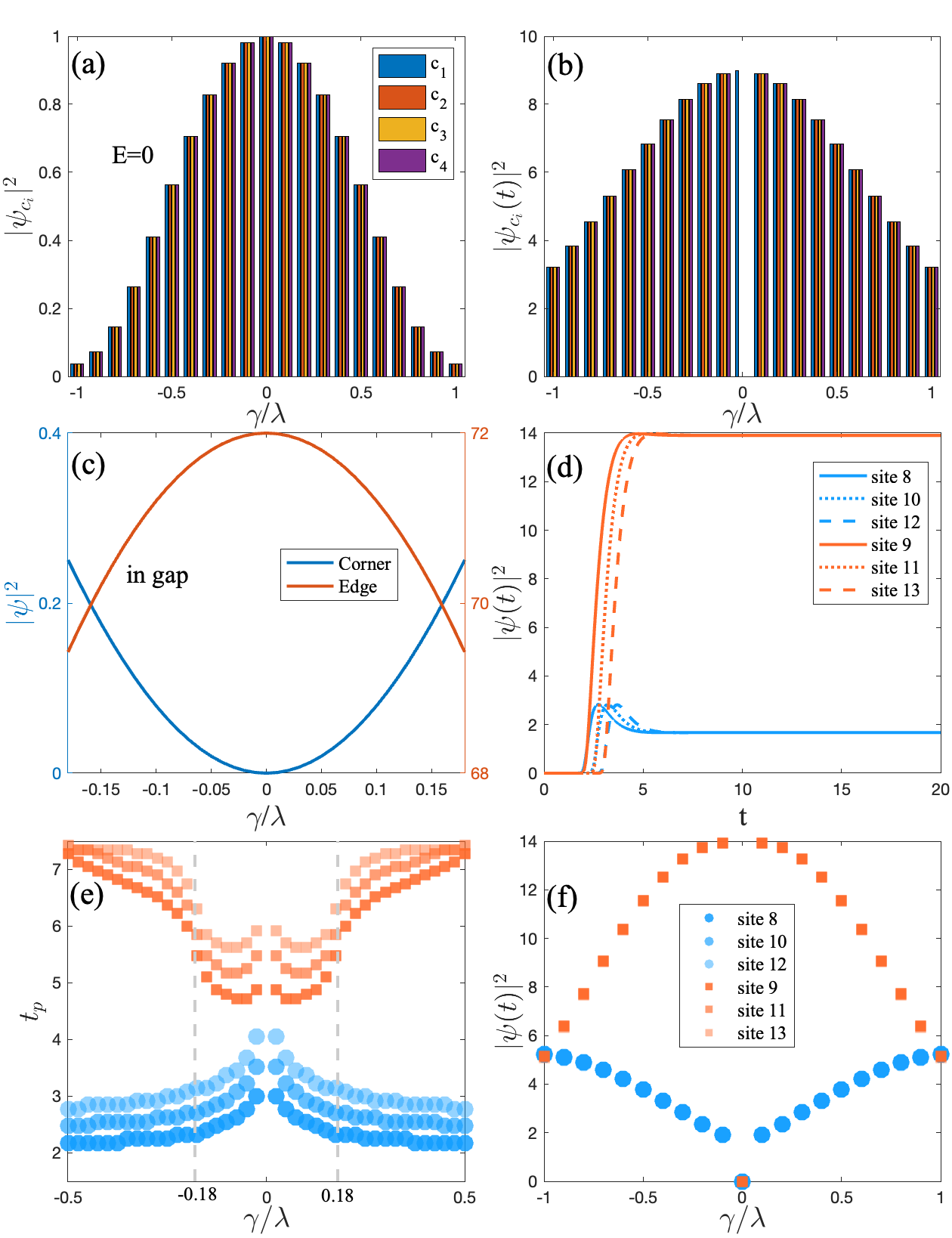}
\caption{ Intensity of four degenerate zero-energy corner states as a function of the parameter ratio $\gamma/\lambda$.(a) for unstimulated case and (b) for stimulated case with long enough evolution time. (c) Intensity of in-gap states on corners (blue) and edge sites (orange) for unstimulated case. Here we employ a double y‑axis. (d) The time evolution of the middle sites on the left side with gain across all non-corner edge sites. (e) $t_{p}$ distributions for the middle sites on the left side. (f) is the final distributions of the middle sites after long time evolution.
}
\label{BBH-histogram}
\end{figure}

Fig.~\ref{BBH-histogram}(a) displays the intensity distribution of degenerate zero-energy states at four corner sites, which exhibits Gaussian-curve like confinement in the parameter space via the tuning of $\gamma/\lambda$. For comparison, (b) shows the final stable distribution of the corresponding four corners in the stimulated steady state, where optical gain is applied uniformly to all corner sites.
When intra-hopping $\gamma$ equals zero, i.e. $\gamma/\lambda=0$, the initially excited corner state $c_1$ fails to transfer the signal to other lattice sites, leading to a final intensity distribution that remains exclusively localized at that corner. In contrast to the static distribution shown in (a), the time-evolved distribution in (b) exhibits considerable occupation across the entire parameter space with similar Gaussian-type shape and remains appreciable at the topological phase transition point $\gamma/\lambda=1$ $(N \rightarrow \infty)$. This is because after time evolution, the corners spatial distribution of the state  contains information about the whole system, taking into account the interplay between corner and edge states, rather than originating solely from either one. However, overall, the topologically nontrivial characteristics still dominate.

For the non-trivial edge states, i.e. in gap states (marked up with green color in Fig.~\ref{BBH-sketch}(b)), we analyze the static specific intensity distribution at corner and edge sites within range $|\gamma/\lambda|<0.18$ in Figure~\ref{BBH-histogram}(c). As expect, the edge component (excluding the corner sites) is the dominant contribution throughout this interval, and it gradually decreases as $|\gamma/\lambda|$ increases. In comparison, the corner component is small but not zero and it increases with increasing $|\gamma/\lambda|$. 
\begin{figure}[h]
\centering
\includegraphics[width=1\linewidth]{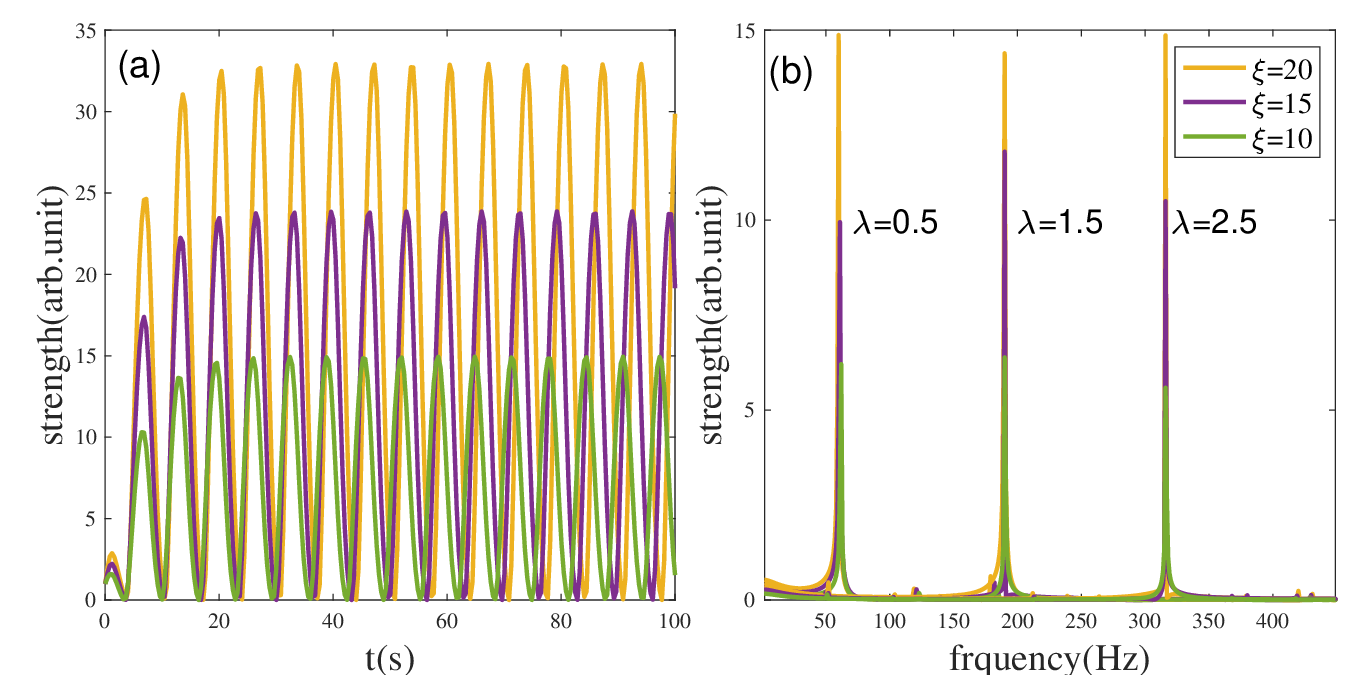}
\caption{(a) Oscillations on the $e_{1}$ site (illustrated in Fig.\ref{BBH-sketch}(a)) with $\gamma=0$, that $\xi$ takes 20, 15, 10. (b) Oscillation strength in the frequency domain while the inter-hopping strength takes $\lambda=0.5, 1.5, 2.5$ respectively. }
\label{oscillation}
\end{figure}

In the finite parameter region $|\gamma/\lambda|<0.18$, topologically protected corner and edge states coexist. To understand how the laser dynamics behaves under their combined interplay and investigate how the in-gap states affect edge dynamics, we apply optical gain exclusively to the lattice edge sites while excluding the four corner sites.  We label the sites along the left edge from $c_{1}$ to $c_{2}$ as $1$-$20$, where the odd- and even-indexed sites correspond to the A- and B-sublattices (analog to 1D SSH), respectively. Specifically, we pump the $c_{1}$ corner and monitor the temporal evolution of the intensity on the sites (No.8-No.13) located in the middle of the left edge in Figure~\ref{BBH-histogram}(d).  We find that the intensities on the A- and B-sublattices each become stabilized to approximately equal excitation intensities. However, since the $c_{1}$ corner (No.1) is driven out of equilibrium, the odd-indexed sites experience stronger stimulation, i.e. $\vert \psi_A(t)\vert^2 > \vert \psi_B(t)\vert^2$. In addition, for each site, the time at which the intensity reaches its extremum during the dynamics scales directly with the site’s distance from $c_1$.
We denote this characteristic time as $t_{p}$. It exhibits a distinct dependence on $|\gamma/\lambda|$, within the parameter range $|\gamma/\lambda|<0.18$, where the edge states exist, as shown in Fig.~\ref{BBH-histogram}(e). The intensity on the B-sublattices reaches its peak earlier than that on the A-sublattices, which is because the A-sublattice intensity is comparatively stronger and it requires a longer time to build up. In particular, for the B-sublattices, $t_{p}$ decreases rapidly as $|\gamma/\lambda|$ increases and becomes approximately constant once the system enters the trivial phase where the edge states merge into the bulk continuous spectrum and disappear. This behavior can be understood as follows: when the edge states are present, the transmission efficiency increases with the larger intra-hopping strength.
Conversely, for the A-sublattices, $t_{p}$ remains nearly constant within the same range $|\gamma/\lambda|<0.18$, but it increases rapidly when the edge states vanish. This indicates that $t_{p}$ is relatively insensitive to the intra-hopping strength while the edge states exist, whereas it becomes significantly larger after their disappearance, i.e., the peak intensity is reached much more slowly. Finally, Fig.~\ref{BBH-histogram}(f) presents the long-time intensity distribution of the A- and 
B-sublattices at the middle of the left edge. It demonstrates the relative advantage of the 
A-sublattice intensity caused by the nonequilibrium stimulation on $c_{1}$ and it becomes weaker as $|\gamma/\lambda|$ increases until the intensities on the A- and B-sublattices gradually converge to nearly equal values.

\begin{figure*}[htbp] 
    \centering
    \includegraphics[width=\textwidth]{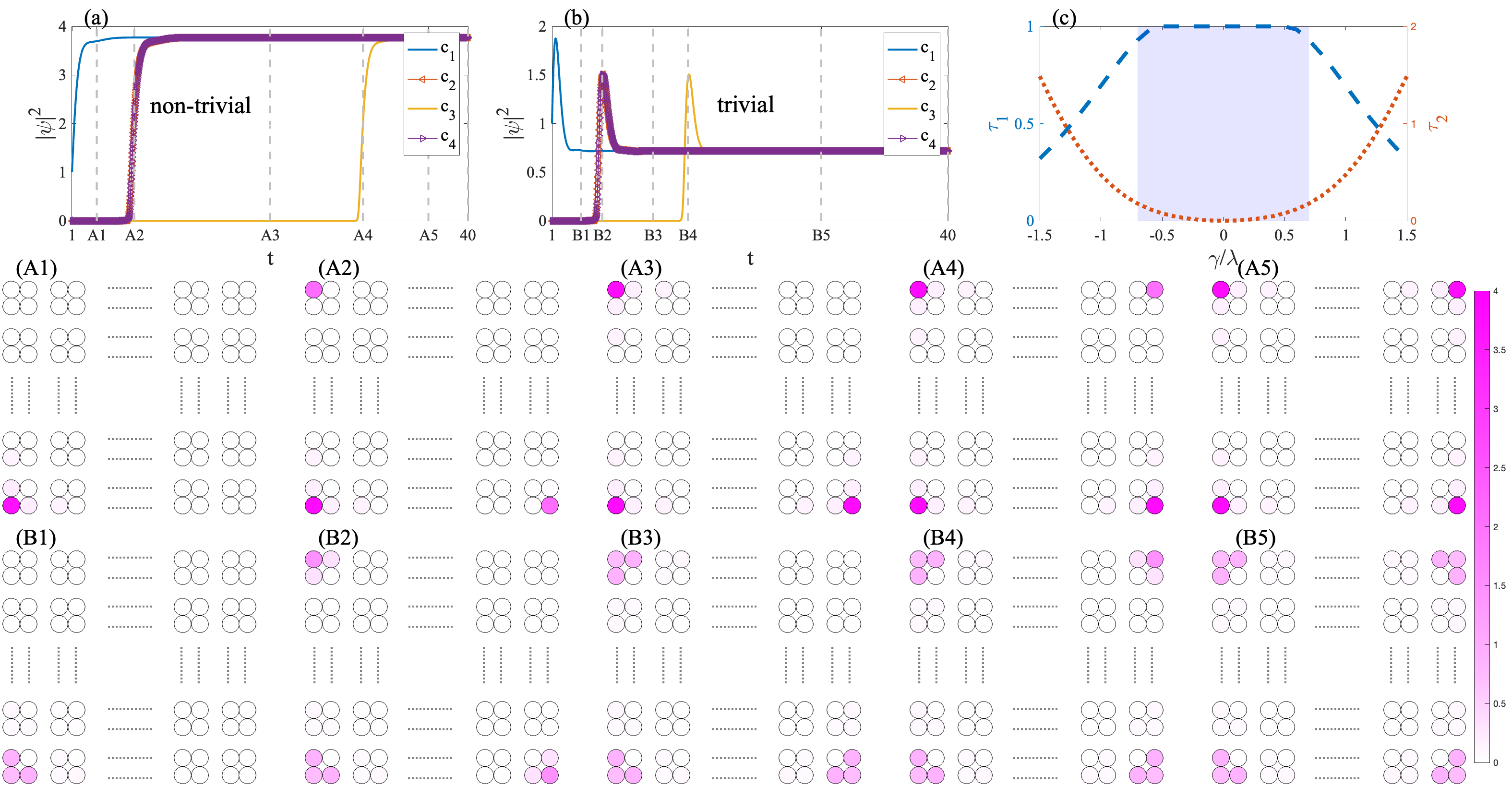} 
    \caption{(a) and (b) dynamics of intensity $|\psi|^{2}$ in the four corners with $\gamma=0.4$ and $\gamma=1.4$.  (c) $\tau_{1}:$ ratio of the stable strength versus maximum strength at corner $c_{1}$. $\tau_{2}:$ ratio of the stable strength at $e_{1}$ versus strength at $c_{1}$. The purple-shaded background region demarcates the parameter range where corner states exist. (A1)-(A5) depict the excitation configurations of the lattice at several specific time points for topological non-trivial phase. The corresponding time for each figure is indicated on the timeline provided in panel (a).   (B1)-(B5) evolution of the lattice at specific time labeled in (b) for topological trivial phase.}
    \label{even-evolution}
\end{figure*}

By analyzing the dynamics of corner-localized states and their mutual transformation within the lattice, we aim to elucidate the distinctions between trivial and non-trivial regions in terms of excitation patterns and decay rates. Here we initially place the gain on the four corners and stimulate $c_{1}$. The time evolutions of corners' intensity behave differently in the non-trivial and trivial regions, see in Fig.~\ref{even-evolution}(a) and (b). In both cases, corners $c_2$ and $c_4$, which are spatially closer to $c_1$, saturate faster than the more remote corner $c_3$. The key signature distinguishing the two topologically distinct dynamics lies in the intensity evolution: in the non-trivial case, intensities grow not only rapidly but also monotonically‌, whereas in the trivial case, the curves decline immediately after a peak labeled as $I_{m}$ and eventually stabilize at a stable value $I_{s}$. 

We  quantitatively  characterize these dynamics using two specific ratios that quantify the system’s behavior. The first ratio, $\tau_{1}=I_{s,1}/I_{m,1}$, quantifies the decay amplitude at site $c_1$, where $I_{s,1}$ and $I_{m,1}$ denote the stable and peak intensity components of this corner, respectively. Notably, $\tau_{1}$ exhibits a distinct change in slope at the topological phase transition, as shown in Fig.~\ref{even-evolution}(c). Within the topologically nontrivial region (shaded in purple), $\tau_{1}$ remains constant at 1, but it begins to decay immediately upon entering the trivial regime. This sharp contrast serves as a clear diagnostic for distinguishing between the two phases.
The second ratio is defined as $\tau_{2}=I_{n,1}/I_{s,1}$ where $I_{n,1}$ denotes the stable intensity at edge site $e_{1}$ (one of the two nearest neighbors of $c_1$ with equal intensity), as shown in Fig.~\ref{BBH-sketch}(a). $\tau_{2}$ reflects the diffusion capacity along the edge and it experiences an accelerated growth approaching the  topological trivial region. Phenomenologically, both ratios exhibit an abrupt change with the tuning of the parameter $\gamma/\lambda$, revealing an intrinsic modulation of the laser output stability by the topology.

Panels (A1)–(A5) and (B1)–(B5) in Fig.~\ref{even-evolution} provide a comprehensive view of signal propagation from initially stimulated site $c_{1}$ across all sites as a function of time, corresponding to the topological non-trivial and trivial dynamics in panels (a) and (b), respectively. In both cases, excitation spreads from $c_{1}$ to nearest neighboring corners $c_{2}$ and $c_{4}$ first, then reaching $c_{3}$ and eventually all the corners' intensity are stabilized at a common value protected by reflection symmetries in the two dimensions. The key distinction emerges in the trivial phase: corner intensities initially peak and subsequently diffuse to neighboring sites. Specifically, $c_{2}$ and $c_{4}$ exhibit peak intensities in (B2) that decay and spread to their neighbor sites in (B3), while $c_{3}$ reaches its maximum in (B4) and disperses in (B5). Comparing A and B panels, we conclude that the topologically nontrivial phase exhibits more localized and intense excitations—a hallmark of the robust corner state. In contrast, excitations in the trivial phase are relatively diffuse and inefficient.

Alternatively, in the isolated corner limit $\gamma=0$, exciting one edge $e_{1}$ site could induce single-frequency oscillations. We endeavor to modulate the amplitude and frequency of the oscillations by manipulating the nonlinear terms in the lasing dynamics. $\alpha\xi$ in the dynamics (Eq. (\ref{Eq1})) represents the stimulation gain and Fig.~\ref{oscillation}(a) exhibits the amplitude strength linearly depends on the varying $\xi$, while $\alpha$ is fixed. In Fig.~\ref{oscillation}(b), we varied the inter-hopping strength $\lambda$ and observed a linear shift in the oscillation frequency after Fourier transformation. The tunability enables designers to engineer laser systems that produce specific amplitudes and frequencies on demand. Unlike conventional frequency-stabilization approaches that rely on external feedback mechanisms or complex mode-locking schemes, the topologically protected nature of these oscillations provides intrinsic robustness. 

 \subsection{Lasing behavior in odd-sized  2D BBH lattice and topological phase transitions}
 \begin{figure}[htbp]
\centering
\includegraphics[width=1\linewidth]{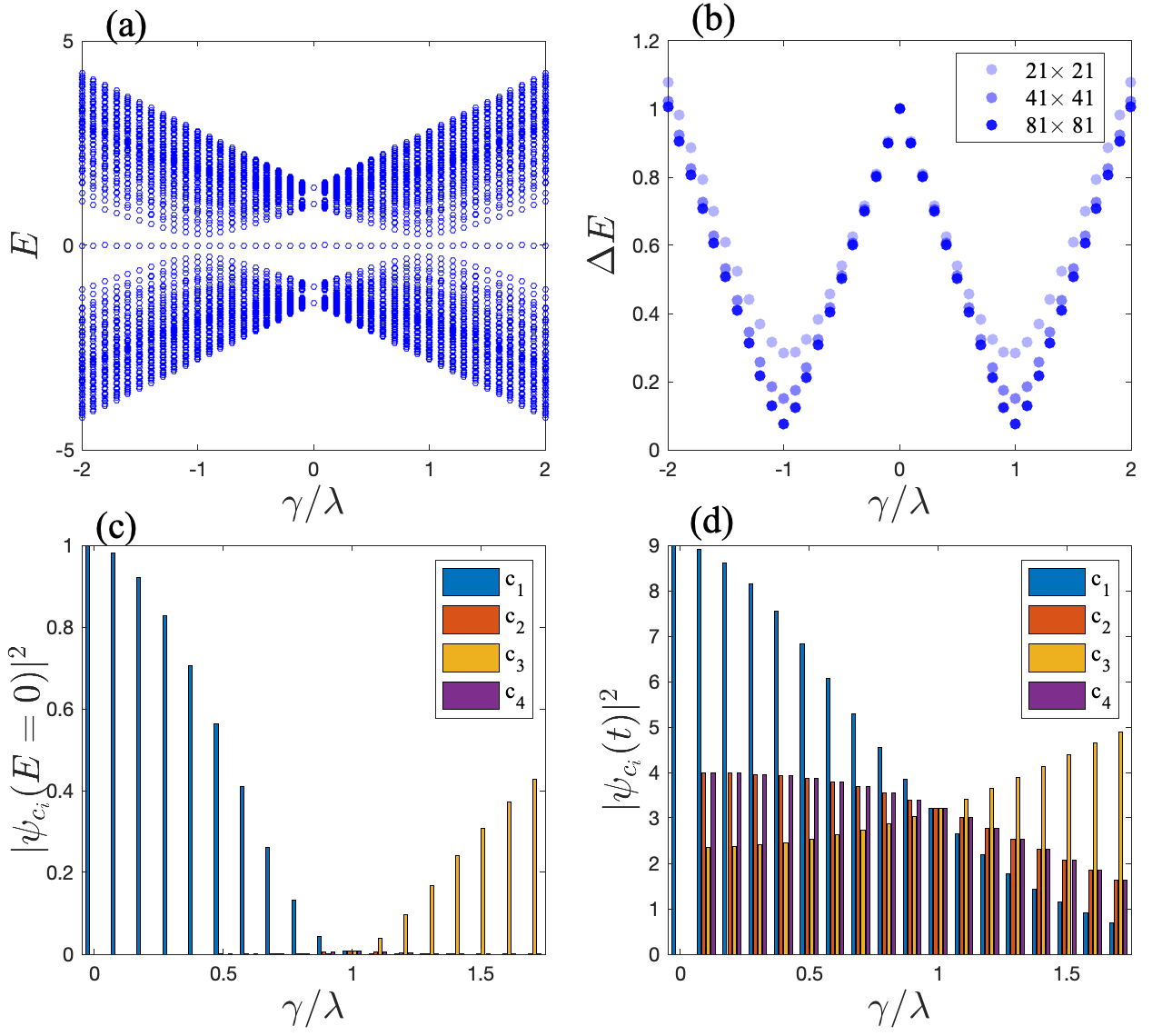}
\caption{(a) The energy spectrum of $21\times21$ BBH model. 
(b) $\Delta E$ refers to the energy gap between $E=0$ and its nearest upper band in $N\times N$ lattices, for which $N=21$, $41$ and $81$ respectively. (c) is the intensity distributions on corners for zero-energy states. (d) is the stabilized distribution on corners after long enough time evolution.}
\label{BBH-odd}
\end{figure} 
\begin{figure*}[htbp]
\centering
\includegraphics[width=\textwidth]{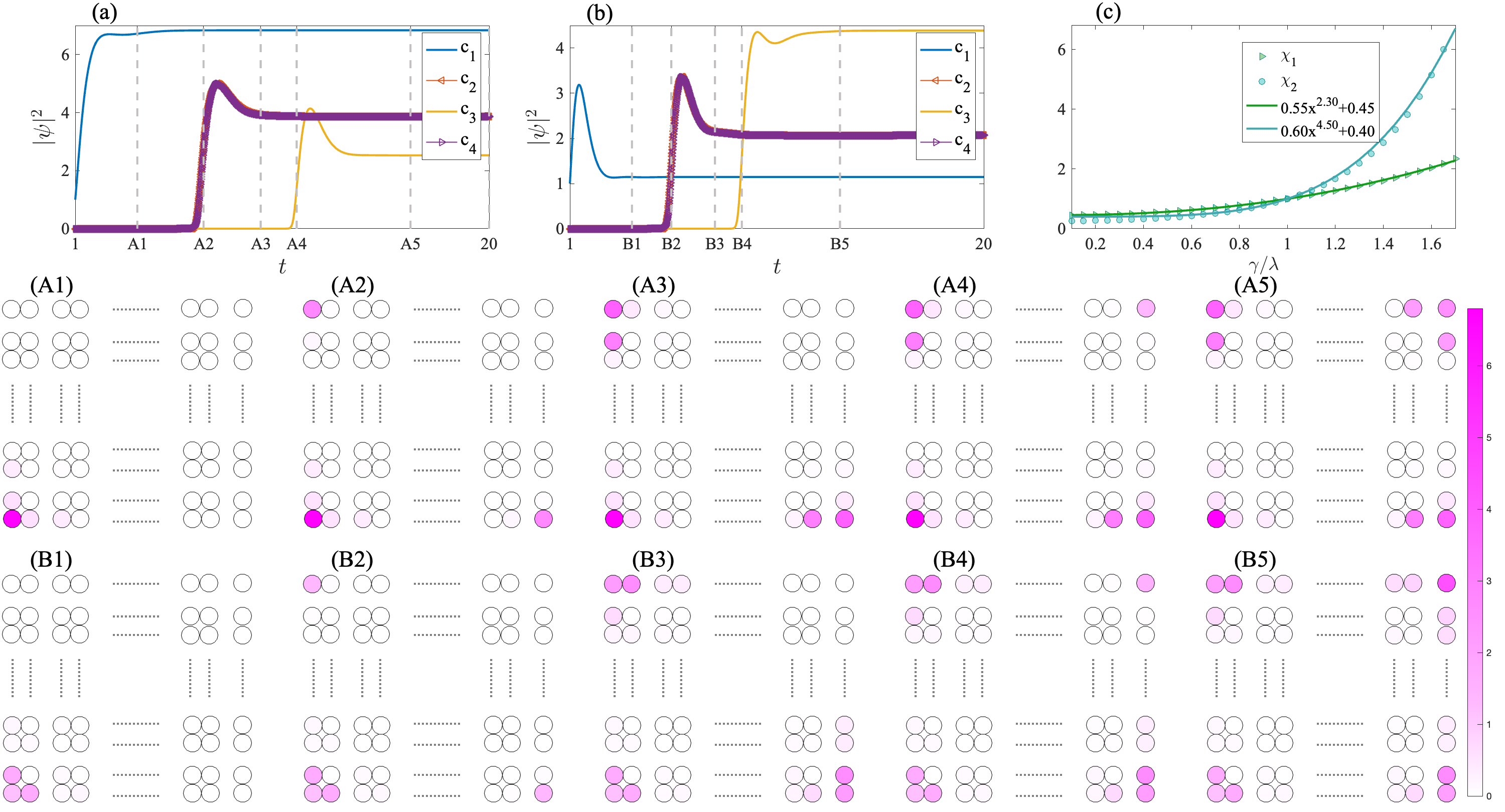}
\caption{(a) and (b) dynamics of intensity $|\psi|^{2}$ in the four corners with $\gamma=0.5$ and $\gamma=1.5$.  (c) $\chi_{1}$, $\chi_{2}$ and their fitting curves. (A1)-(A5) evolution of the lattice at specific time as labeled in (a). (B1)-(B5) evolution of the lattice at specific time labeled in (b). }
\label{odd-evolution}
\end{figure*}
The $N \times N$ BBH lattice with odd $N$ exhibits distinct topological phases in its energy spectrum and characteristic localization signatures of zero-energy states that differ markedly from the even-$N$ case, as shown in Fig. \ref{BBH-odd}(a). Notably, only when $|\gamma/\lambda=1|$, the zero-energy state has multi-fold degeneracy, elsewhere it is a single zero-energy state. To assess the robustness of these features against finite-size effects, we examine the energy gap between the zero-energy band and its nearest-neighboring band. We find that this gap $\Delta E$  reaches its minimum at the point ($|\gamma/\lambda=1|$), which is the phase transition point and decreases monotonically with increasing system size in (b). 

Figure~\ref{BBH-odd}(c) illustrates the corners' intensity distribution of zero-energy states as a function of $\gamma/\lambda$. For $|\gamma/\lambda<1|$, particularly when $\gamma/\lambda=0$, the intensity is localized solely at site $c_{1}$. As $|\gamma/\lambda|$ increases toward 1, the intensity gradually delocalizes and spreads across edge and bulk sites, eventually becoming uniformly distributed across the A-sublattice and maintaining very small value. When $|\gamma/\lambda|$ exceeds 1 and continues to increase, the uniform distribution progressively breaks and begin to concentrate at corner $c_{3}$.

The odd-numbered lattice features A-sublattice terminations at all four corners. The zero-energy state solutions for the amplitude are given by \cite{BergholtzPRB18}: 
\begin{equation}
  \vert \Psi \rangle = \mathcal{N} \sum_{m=1}^N \sum_{m'=1}^N r^{(m+m')/2} c_{A,m,m'}^\dagger \vert 0 \rangle 
\end{equation}
where $m$ and $m'$ are all odd numbers in the $N\times N$ BBH lattice. Here, $\mathcal{N}$ is the normalized term,  and $r = -\gamma/\lambda$, which implies that the state is exponentially localized at the bottom-left corner $c_1$ when $\gamma/\lambda < 1 $, and at the top-right corner ($c_3$) when $\gamma/\lambda > 1$, as depicted in Fig.~\ref{BBH-odd}(c). The distinct corner state patterns observed in this system, compared to an even-N system, arise from the breaking of reflection symmetry. Since one edge of the BBH lattice shares the same configuration as the 
1D SSH chain, we examine the lasing dynamics of SSH chains with even and 
odd numbers of sites (Appendix~\ref{app:1D SSH}), including those with a central defect (Appendix~\ref{app:1D SSH_defect}), to clarify 
the correspondence between dynamical signatures and topological phases.

We apply optical gain to all four corners and selectively stimulate the $c_{1}$ corner in Fig.~\ref{BBH-odd}(d). At the isolated corner limit $\gamma/\lambda=0$, excitation remains confined to the initially stimulated corner $c_{1}$. For $|\gamma/\lambda|>0$, all four corners become excited with intensity hierarchy $c_{1}>c_{2}=c_{4}>c_{3}$, which contrasts with the intensity distribution arising solely from zero-energy states in (c). The equality $c_{2}=c_{4}$ is protected by the inversion symmetry. With increasing $|\gamma/\lambda|$, the intensities at $c_{1}$, $c_{2}$ and $c_{4}$ monotonically decrease while the intensity at $c_{3}$ grows; these curves intersect at $|\gamma/\lambda=1|$. At larger $|\gamma/\lambda|$, $c_{3}$ becomes the dominant excitation site. This progressive shift in spatial distribution directly reflects the topological phase transition occurring at $|\gamma/\lambda=1|$.

Panels (A1)–(A5) and (B1)–(B5) in Fig.~\ref{odd-evolution} present a comprehensive spatiotemporal view of signal propagation originating from the initially stimulated site $c_{1}$ across all corner sites as a function of time, contrasting two distinct topological phases shown in panels (a) and (b), respectively. In both cases, excitation propagates from $c_{1}$ to the nearest neighboring corners $c_{2}$ and $c_{4}$ before reaching the diagonally opposite corner $c_{3}$. However, the final intensity distributions differ markedly depending on the parameter $\gamma/\lambda$: specifically,  $c_{1}>c_{2}=c_{4}>c_{3}$ for $\gamma/\lambda=0.5$ and $c_{3}>c_{2}=c_{4}>c_{1}$ for $\gamma/\lambda=1.5$. This observation reflects the distinct dynamical signatures present in different topological phases. 

To quantify the corner intensity relationships, we introduce the ratios
$\chi_{1}=I_{c_2}/I_{c_1}$ and  $\chi_{2}=I_{c_3}/I_{c_1}$, where 
$I_{c_i}$ denotes the steady-state intensity at corner $i$ following sufficient time evolution. Curve fitting these ratios as functions of the parameter $\gamma/\lambda$ in Fig.~\ref{odd-evolution} (c) yields: $y=0.55(\gamma/\lambda)^{2.3}+0.45$ for $\chi_{1}$ and $y=0.60(\gamma/\lambda)^{4.5}+0.40$ for $\chi_{2}$. Notably, the exponent for $\chi_{2}$ (4.5) is approximately twice that for
$\chi_{1}$ (2.3). This relationship directly reflects the difference in propagation path lengths along the edge: the distance from $c_{1}$ to $c_{3}$ is approximately twice that from $c_{1}$  to $c_{2}$. It demonstrates that intensity loss is cumulative and geometrically organized, not random. Also, this non-linear scaling reflects the underlying topological structure and indicates that the system exhibits coherent edge-state dynamics where dissipation effects are amplified over longer paths. Thus, $\chi$ serves as a diagnostic tool for characterizing both the spatial localization properties and edge-state quality of the topological system, with the power-law dependence on $\gamma/\lambda$ encoding how the balance between coupling strength and dissipation determines corner-to-corner connectivity.

\section{Conclusion}
In this study we systematically investigate the two-dimensional BBH model as a platform for higher-order topological nano-lasers, elucidating the profound influence of topological properties on lasing dynamics and performance. Our findings underscore the intrinsic advantages of leveraging higher-order topology for robust, efficient and tunable light sources. We have rigorously demonstrated that the lasing behavior is fundamentally governed by the topological attributes of the BBH Hamiltonian. Critically, the final stable light excitation, after prolonged evolution, is largely determined by these topological invariants, leading to emission profiles intrinsically linked to the localized corner states. This topological dictation of emission provides a paradigm shift from conventional laser design, where complex cavity engineering is often required to achieve stable single-mode operation.

Our introduction of two key diagnostic observables $\tau_{1}$ and $\tau_{2}$, provides a quantitative framework for characterizing topological phase transitions dynamically. $\tau_{1}$, representing the ratio of saturate to peak intensity at a corner site, exhibits a distinct and abrupt change at the topological phase boundary ($|\gamma/\lambda|=1$), serving as a clear experimental signature distinguishing nontrivial from trivial phases. Furthermore, $\tau_{2}$, which quantifies edge diffusion, offers insights into the delocalization processes associated with topological phase transitions. The consistency of these indicators across different lattice configurations reinforces their utility as robust probes for topological characterization.

We further elucidate the role of lattice parity in modulating corner state localization. 
In odd-sized lattices, corner states exhibit a corner-selective localization, appearing exclusively on either $c_{1}$ or $c_{3}$ sites. This selectivity is tuned by the ratio $|\gamma/\lambda|$, such that a topological transition, driven by specific symmetry-breaking mechanisms, occurs as this parameter crosses the critical value of 1. This parity-dependent behavior offers an additional degree of freedom for engineering the spatial distribution of topological modes. 

Moreover, the quantitative relationship between the exponents of intensity ratios ($\chi_{2}\approx 2\chi_{1}$) directly correlates with the geometric propagation distances along the lattice edges, establishing a clear link between topological dynamics and physical geometry.
 Beyond their inherent robustness to defects, the topological corner states in the BBH model confer superior lasing characteristics when compared to edge-state-based systems. These include enhanced spatial confinement, reduced effective mode volume, and significantly higher quality factors, all of which contribute to lower lasing thresholds and improved efficiency. The demonstration of topologically protected oscillations in the isolated corner limit, with tunable amplitude and frequency via nonlinear gain parameters, presents a novel avenue for on-demand frequency stabilization without reliance on external feedback.
 
In conclusion, this work significantly advances the understanding of topological dynamics in active photonic systems. By thoroughly exploring the BBH model, we have not only revealed the fundamental mechanisms underlying topological corner-state lasing but also established clear pathways for the design and optimization of next-generation topological photonic devices. While the precise control of bistability between corner and edge states remains an intriguing challenge, the foundational insights provided herein pave the way for future innovations in compact, robust, and efficient nano-lasers, integrated photonic circuits, and advanced topological quantum optical systems.

\section{Acknowledgments}
ZYL was supported by the National Natural Science Foundation of China (Grant Nos.12505045 and 12247101), and by the Natural Science Foundation of Gansu Province (Grant No.25JRRA799). QL was supported by National Natural Science Foundation of China under Grant No.12504174. ZXH  was supported by National Natural Science Foundation of China under Grant Nos. 12474140  and 12547101.

\appendix
\section{Impact of lattice site parity (even vs. odd) on corner states and lasing dynamics in 1D SSH chains}
\label{app:1D SSH}
This appendix provides additional examples about lasing dynamics in 1D SSH lattices. We delves into the energy spectrum and lasing dynamics of finite SSH chains, focusing on how varying parameters, such as the ratio of intracell to intercell couplings ($\gamma/\lambda$), influence the localization and propagation of excitation. By investigating chains with both even and odd numbers of sites, we uncover essential insights into the impact of lattice structure on topological properties and the corresponding behavior of lasing dynamics. 

Within the tight-binding framework, the SSH Hamiltonian can be expressed in the following form:

\begin{equation}
H_{SSH}=\sum_{n}\gamma\hat{c}^{\dag}_{A,n}\hat{c}_{B,n}+\lambda\hat{c}^{\dag}_{A,n+1}\hat{c}_{B,n}+h.c.
\end{equation}

where $\hat{c}^{\dag}_{A,n}$ and $\hat{c}^{\dag}_{B,n}$ denote photon creation operators at site $n$ in the sublattices A and B of this one-dimensional chain, $\gamma$ and $\lambda$ are the intracell and intercell coupling coefficients, respectively.

\begin{figure}[htpb]
\centering
\includegraphics[width=1\linewidth]{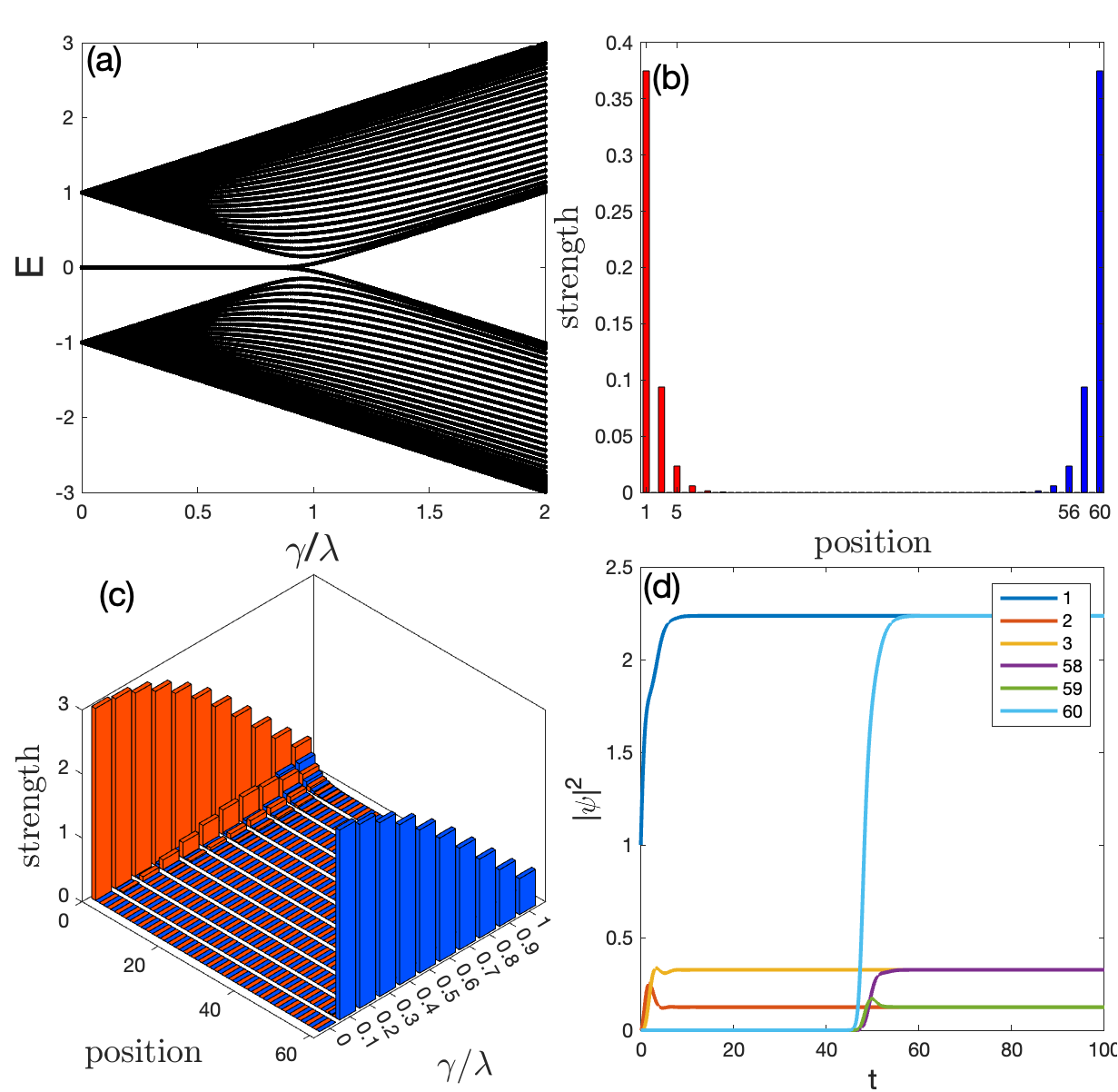}
\caption{(a) The energy spectrum of the finite 1D SSH model with 60 sites. (b) The localized corner states at $E=0$ with $\gamma/\lambda=0.6$. (c) The final stationary strength of evolving states via a stimulated emission on the 1st site versus $\gamma/\lambda$. (d) Intensity versus t for various sites when $\gamma/\lambda = 0.6$.
}
\label{SSH-even}
\end{figure}
Fig.~\ref{SSH-even}(a) displays the energy spectrums of a finite SSH chain with even sites. Two topologically distinct phase are separated in parameter space, where $\gamma/\lambda < 1$ denotes the topological non-trivial phase with zero-energy corner state  and  $\gamma/\lambda > 1$ is trivial phase.  In  topologically nontrivial phase, the chiral symmetry gives rise to a pair of zero-energy edge modes localized at opposite ends and on distinct sublattices, constituting a two-fold degeneracy (shown in Fig.~\ref{SSH-even}(b)).
Based on its fundamental topological properties and relative basic structure, we firstly put gain on two ends of the chain and initially stimulated the leftmost site. 
Since parameter ratio $\gamma/\lambda$ controls the topological phase, we track the occupation or intensity of each sites of the chain after enough evolution time in Fig.~\ref{SSH-even}(c). When intracell hopping $\gamma=0$,   the stimulated photon is localized at 1st sites with other sites unexcited.  Gradually increasing the parameter $\gamma/\lambda$ leads to significant excitations of equal intensity at both ends of the chain. However, as $\gamma/\lambda$ approaches 1, the energy gap between the upper and lower bands  gradually diminishes, and the system begins to enter a trivial phase. Consequently, the excitation characteristics at two ends become less distinct compared to other lattice sites.
The intensity $|\psi|^{2}$
  at the sites around the two ends with $\gamma/\lambda=0.6$ is shown in Fig.~\ref{SSH-even}(d). Here, the timing of stimulation turning on depends on the distance from the initial site, and only the two very end sites maintain significant intensity, in contrast to the system's topological profile. 

\begin{figure}[htbp]
\centering
\includegraphics[width=1\linewidth]{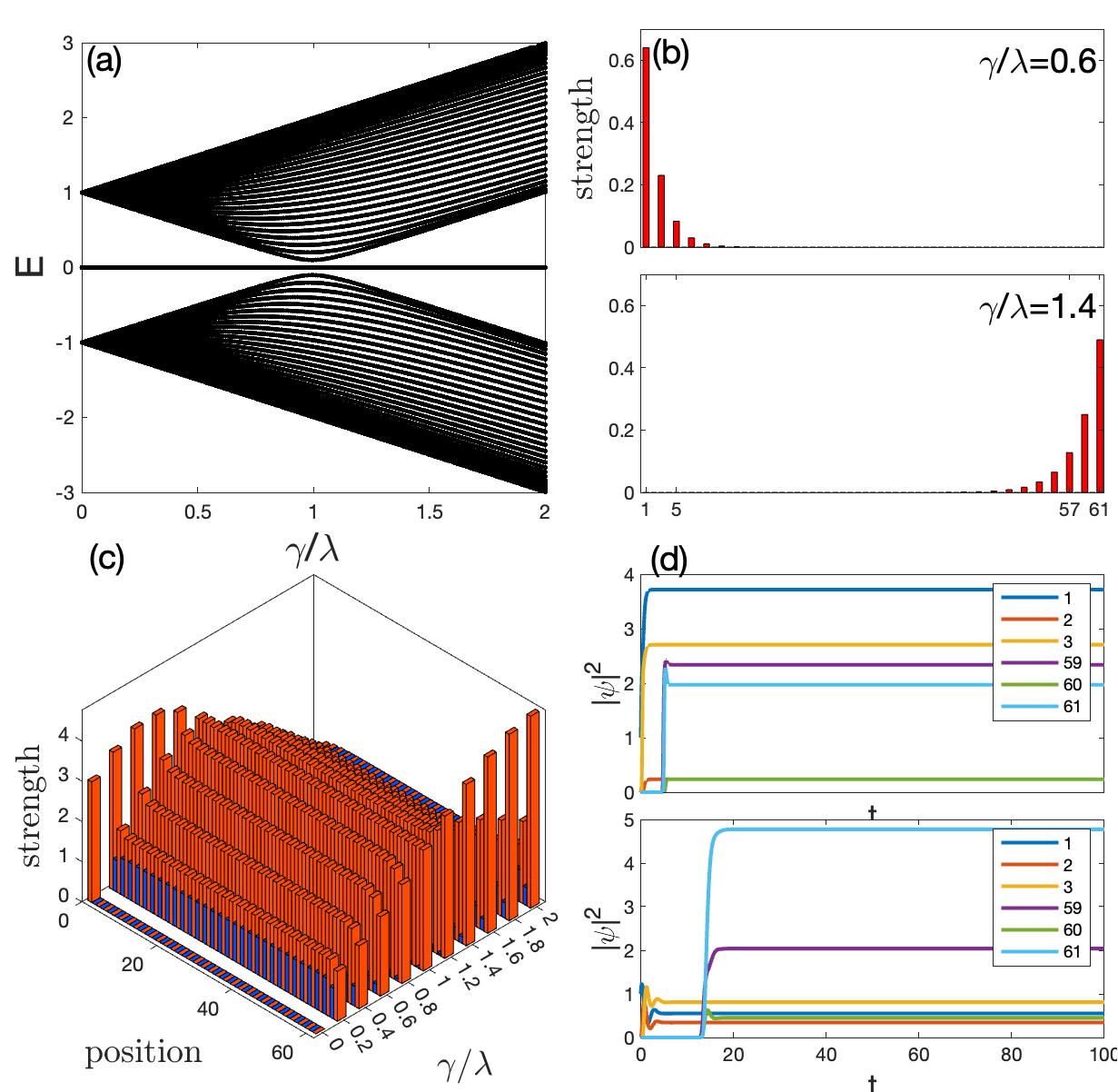}
\caption{(a) The energy spectrum in the finite 1D SSH chain model with 61 sites. (b) The localized corner states at $E=0$ while upper and lower panels refer to $\gamma/\lambda=0.6, 1.4$. (c) The final stationary strength of evolving states via a stimulated emission on the 1st site while $\gamma/\lambda$ takes different values.
(d) Strength changes with time at the two ends of the chain that upper and lower panels refer to  $\gamma/\lambda=0.6, 1.4$ respectively.}
\label{SSH-odd}
\end{figure}

 For finite SSH chains with odd number of sites, the topological non-trivial phase exists in both $\gamma/\lambda<1$ and $\gamma/\lambda>1$ regimes, while the gap energy reaching its minimum at $\gamma/\lambda=1$ (Fig.~\ref{SSH-odd}(a)). We analyze a pair of edge-localized in-gap states with symmetric parameters about $\gamma/\lambda=1$ in Fig.~\ref{SSH-odd}(b).
The odd number lattice has A-sublattice terminations at two ends. This specific boundary condition is crucial for rendering the edge states analytically solvable. For SSH model, an exact end state could be given as follows~\cite{BergholtzPRB18}:
 \begin{equation}
     \vert \Psi \rangle = \mathcal{N} \sum_{m=1}^M  r^m c_{A,m}^\dagger \vert 0 \rangle
 \end{equation}
 where $m$ spans all lattice sites with odd integer coordinates and $r = -\gamma/\lambda$ implying the state is exponentially localized at the left($\gamma/\lambda <1$) or to the right when $\gamma/\lambda > 1$, as seen in Fig.~\ref{SSH-odd}(b).
 
To elucidate the role of odd-site localization in the system dynamics, we apply uniform gain to all odd sites and inject the excitation at the leftmost boundary. The resulting spatial intensity distribution with the corresponding parameters is presented in Fig.~\ref{SSH-odd}(c). 
For the isolated corner limit $\gamma/\lambda=0$, the intensity remains solely at the corner. As the parameter increases, the intensity at the first site decreases while that at the opposite end increases. Meanwhile, the intensity at non-corner odd sites grows and get equivalent until $\gamma/\lambda=1$ with the even sites are suppressed. For $\gamma/\lambda>1$, the trend at the two end sites continues, but the intensity at non-corner sites declines.
In Fig.~\ref{SSH-odd}(d), we plot the intensity evolution over time near the two ends. More odd sites are stimulated for $\gamma/\lambda<1$, comparing to the case for $\gamma/\lambda>1$.  

\section{Effect of central defects on lasing dynamics in  1D SSH model}
\label{app:1D SSH_defect}
\begin{figure}[htbp]
\centering
\includegraphics[width=1\linewidth]{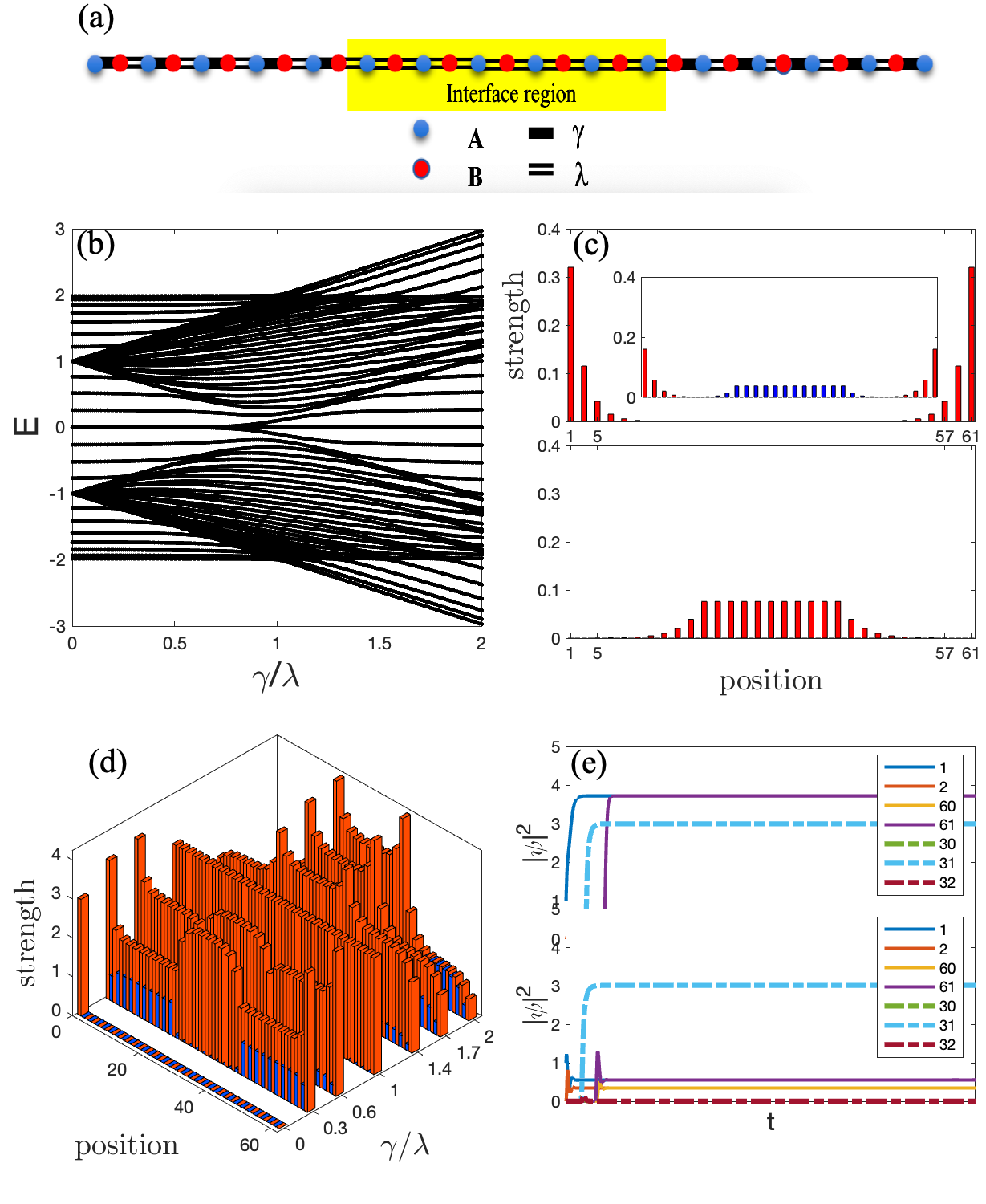}
\caption{(a) Sketch of SSH model with central defect. (b) The energy spectrum in the finite 1-D SSH chain model with central defect region. (c) The localized corner states at $E=0$ while upper and lower figures refer to $\gamma/\lambda=0.6, 1.4$.
(d) The final stationary strength of evolving states via a stimulated emission on the 1st site while $\gamma/\lambda$ takes different values. (e) Strength changes with time at the two ends and defect region of the chain, where the upper and lower panels refer to  $\gamma/\lambda=0.6, 1.4$ respectively.}
\label{SSH-defect}
\end{figure}

We introduce a central defect region where the hopping strengths between all neighboring sites are set equal to the inter‑hopping strength of the non‑defect region, so as to preserve mirror symmetry across the entire chain (see Fig.~\ref{SSH-defect}(a)).
The energy spectrum in Fig.~\ref{SSH-defect}(b) exhibits a three-fold degeneracy in the zero-energy states for $\gamma/\lambda<1$, while this degeneracy is lifted and only a single  zero-energy state remains for $\gamma/\lambda>1$.
Among the three degenerate states for $\gamma/\lambda<1$, one is an edge mode localized at the chain ends, while the other two are hybridized states whose wavefunction densities are distributed both at the ends and across the central defect region.  For $\gamma/\lambda>1$, the single non-trivial state shows localization exclusively on the defect region (Fig.~\ref{SSH-defect}(c)). We apply gain to every odd site and stimulate the first site. After sufficient evolution time in Fig.~\ref{SSH-defect}(d), for $\gamma/\lambda<1$, we observe strong excitation of all odd sites, particularly at the corner and defect regions, which reflects the non-trivial hybrid profile of the zero-energy states. For $\gamma/\lambda=1$, bulk emission occurs predominantly on the odd sites. For $\gamma/\lambda>1$, the density in the non-defect region exhibits gradual decay. By comparing the dynamics for different topological phases in Fig.~\ref{SSH-defect}(e), we find that corner excitations cannot be stimulated for for $\gamma/\lambda>1$ while the defect region odd sites remain excited. The defect-tuned topological structure offers a platform for reconfigurable topological lasers that dynamically switch between distributed multi-mode and defect-confined single-mode operation by modulating the hopping strength, enabling flexible spatial mode engineering without requiring external dissipation control.

\bibliography{ref}

\end{document}